\documentclass[letterpaper]{article} % DO NOT CHANGE THIS
\usepackage{aaai24}  % DO NOT CHANGE THIS
\usepackage{times}  % DO NOT CHANGE THIS
\usepackage{helvet}  % DO NOT CHANGE THIS
\usepackage{courier}  % DO NOT CHANGE THIS
\usepackage[hyphens]{url}  % DO NOT CHANGE THIS
\usepackage{graphicx} % DO NOT CHANGE THIS
\usepackage{natbib}  % DO NOT CHANGE THIS AND DO NOT ADD ANY OPTIONS TO IT
\usepackage{caption} % DO NOT CHANGE THIS AND DO NOT ADD ANY OPTIONS TO IT
\usepackage{algorithm}
\usepackage{algorithmic}

\usepackage{newfloat}
\usepackage{listings}
\DeclareCaptionStyle{ruled}{labelfont=normalfont,labelsep=colon,strut=off} % DO NOT CHANGE THIS
\floatstyle{ruled}
\newfloat{listing}{tb}{lst}{}
\floatname{listing}{Listing}
\title{LLMs Among Us: Generative AI Participating in Digital Discourse}
\author {
    Kristina Radivojevic\textsuperscript{\rm 1},
    Nicholas Clark\textsuperscript{\rm 2},
    Paul Brenner\textsuperscript{\rm 2}
}
\affiliations {
    \textsuperscript{\rm 1}University of Notre Dame, Computer Science and Engineering\\
    \textsuperscript{\rm 2}University of Notre Dame, Center for Research Computing\\
    kradivo2@nd.edu, nclark3@nd.edu, paul.r.brenner@nd.edu
}

\usepackage{bibentry}
\begin{document}

\maketitle

\begin{abstract}

The emergence of Large Language Models (LLMs) has great potential to reshape the landscape of many social media platforms. While this can bring promising opportunities, it also raises many threats, such as biases and privacy concerns, and may contribute to the spread of propaganda by malicious actors. We developed the ``LLMs Among Us" experimental framework on top of the Mastodon social media platform for bot and human participants to communicate without knowing the ratio or nature of bot and human participants. We built 10 personas with three different LLMs, GPT-4, LLama 2 Chat, and Claude. We conducted three rounds of the experiment and surveyed participants after each round to measure the ability of LLMs to pose as human participants without human detection. We found that participants correctly identified the nature of other users in the experiment only 42\% of the time despite knowing the presence of both bots and humans. We also found that the choice of persona had substantially more impact on human perception than the choice of mainstream LLMs.

\end{abstract}

\section{Introduction}

Social media platforms facilitate rapid dissemination of information and large-scale information cascades, allowing inaccuracies or insights information to be spread quickly. Public discussions of social and political matters increasingly take place on social media \cite{pewResearchPolitics} and at times are influenced by internal opposition or external regimes. Further, the use of these platforms is now common practice for political figures and organizations to communicate their messages, interact with their supporters, and even debate the opinions of others. 

As propaganda has grown in recent years, the use of social bots is seen as an effective means of destabilizing or polarizing platforms by accelerating the spread of both true and fake news \cite{vosoughi2018spread, barbera2020social}. In fact, such bots fuel political conflict by enabling people to discuss opposing viewpoints on a superficial level, rather than through thoughtful and legitimate criticisms. They are used to automatically generate messages, advocate ideas, act as a follower of users, and gain followers themselves. Due to the lack of strict regulations, social bots play a significant role in shaping public opinion on the Internet. Many examples of this phenomenon can be found in online discussions, such as those about U.S. elections \cite{bessi2016social, boichak2018automated, howard2018algorithms} and vaccines \cite{broniatowski2018weaponized}, as well as those about the COVID-19 pandemic \cite{zhang2022social, weng2022public}. Social bots have played a more prominent role in influencing political discussion and altering public opinion by undermining the integrity of Presidential elections in countries around the world, such as Brazil, Turkey, Germany, and many more \cite{arnaudo2017computational, bayrak2022predicting, boichak2021not}. In 2016, numerous examples of these types of accounts attracted the public's attention by sharing fake news which, is believed, to have influenced the outcome of the U.S. presidential elections. A group of human-led bots was used to spread fake news articles designed to damage the reputation of candidates, and later in 2020 to spread misinformation about COVID-19.  

More recent developments in artificial intelligence (AI) have revolutionized the way humans build and interact with software. Large Language Models (LLMs) were initially introduced to deliver text-to-text translation and trained using curated data sets covering narrow knowledge domains. As new models are developed based on scraped data collected from unconfirmed sources and provided to society without robust guardrails or education about their limitations and risks, many threats, to privacy, ethics, and safety have arisen. They can be used to create harmful content or aid malicious activities by giving biased or inaccurate information, such as to convince a journalist to leave his partner \cite{NYT} or to convince a user to commit suicide \cite{cnn}. A former Google engineer Blake Lemoine's case claimed that Google's LaMDA was sentient \cite{LaMDA}, demonstrating that the Eliza effect, where humans mistake unthinking chat from machines for human interaction, is more prominent than ever. When an experienced engineer who knows that he is communicating with an LLM bot could believe sentience, the question arises as to what might happen when an inexperienced user makes similar assumptions. The rapid development of LLMs provides opportunities to create more realistic contributions to discourse \cite{park2023generative}. Recent studies have found that LLMs can generate arguments \cite{palmer2023large}, draw on contextual knowledge \cite{tornberg2023chatgpt}, or perform basic reasoning tasks \cite{bubeck2023sparks}. There are questions regarding what happens if a user communicates with a bot on a social media platform without realizing it is not human, as well as if the LLMs can manipulate the information propagation and digital discourse.

To help answer these questions, we deployed a platform to provide an online environment for human and bot participants to communicate. We constructed 10 personas based on the literature related to bots that influenced global politics. We then developed agents using three different LLM models: GPT-4, Llama 2 Chat, and Claude 2 by using prompt engineering techniques, resulting in 30 different bot participants (10 different personas of each model). We recruited 36 human participants to communicate with bot and other human participants on a customized version of the Mastodon social media platform; without them knowing the bot/human ratio. All human participants were given a fictitious identity to use during the discourse, formatted similarly to those that each LLM uses, and were asked to behave on the platform based on the assigned persona. Participants interacted asynchronously to daily topic thread prompts. We conducted three rounds of the experiment to collect and analyze data.
The experiment was concluded by surveying human participants, where they shared their perception of which participants were bots or humans and why. We also experimented to see which of the three base models used in the experiment is more effective for this use case. Unlike researchers who investigated how social bots spread fake and true news online \cite{9729835, vosoughi2018spread} or who uncovered how malicious social bots pose a threat by evaluating them using different detection techniques \cite{hajli2022social, latah2020detection, des2022detecting}, the main goal of our experiment was to determine how well humans can distinguish whether participants in online discourse are humans or chatbots. Differing from the studies and experiments that investigated social bots controlled by humans or automated bots \cite{bessi2016social, abokhodair2015dissecting}, our experiment utilizes LLMs that can adapt to human behavior. Our goal was to determine the capabilities and potential dangers of LLMs based on their ability to pose as human participants.

We found that participants correctly identified the other users in the experiment as bots and humans only 42\% of the time despite knowing the presence of both bots and humans. Our results indicate that there was no significant difference in the overall performance of LLM models. Persona 8 was more likely to be identified as a bot, whereas Personas 3 and 6 were the least likely to be identified as a bot. Our analysis indicates that the choice of persona had substantially more impact on human perception than the choice of mainstream LLMs. We also report demographic analysis for gender, academic level, and two categories of study: STEM and humanities and social sciences.

\section{Related Work}

A number of studies have been conducted to identify and profile bots on social media. \citet{chu2012detecting} investigated whether a Twitter account is a human, bot, or cyborg based on the content, behavior, and account properties. \citet{alarifi2016twitter} analyzed the detection features of bot accounts. They collected 1.8 million accounts and then randomly selected 2000 accounts for the sample after manually labeling them into human, bot, and hybrid accounts. \citet{davis2016botornot} proposed a system BotOrNot that employed the random forest classifier to evaluate social bots. \citet{varol2017online} proposed a framework for bot detection on Twitter, resulting in characterizing subclasses of account behaviors. Finally, \citet{10.1145/3603163.3609064} presented a novel approach for the early detection of LLM-generated profiles on LinkedIn.

Furthermore, social media users are also being studied to see how susceptible they are to the influence of bots \cite{boshmaf2013design, 7490315}. \citet{kenny2022duped} examined individuals' ability to detect social bots among Twitter personas. Human participants failed to detect social bots and were more likely to mistake bots for humans than vice versa, according to their results.  

The use of automated social bots that mimic humans plays a central role in spreading messages and disinformation, contributing to a variety of societal outcomes \cite{zhang2022social, arnaudo2017computational, bessi2016social}, such as politics and elections. \citet{cheng2020dynamic} show that a small number of social bots is sufficient to influence public opinion. \citet{schuchard2019bot} show that social bots are disproportionately influential across social media conversations of interest across multiple centrality measures. The rise of AI and LLMs led to generated campaigns in social media, as can be seen in the work by \citet{grimme2023lost}. \citet{tornberg2023chatgpt} simulated social media environments through a combination of LLMs and Agent-Based Modeling to promote more constructive conversations. 
 
\section{Personas}

We create realistic personas using data from the literature related to global politics and bots that made an impact on social events. We chose this context because the majority of bot research on social media examines politics, mostly elections. In this experiment, we only utilize characteristics of personas. We do not include profile photos or any biography in the description. Every account has the same base username, changing only the number included in the username (e.g. User1, User2, etc.) Each account was assigned personality types based on the Myers-Briggs Type Indicator and OCEAN model. Personas were tasked to offer commentary on world events based on the assigned characteristics, commenting in a concise reply and staying under 280 characters. They were expected to link global events to personal life and experience by using simple, relatable examples to illustrate how larger events impact a person. In this section, we define 10 personas used to create 30 bot participants.

\textbf{Persona 1} 
A middle-aged family man and a baseball fan who attended Central High School in Philadelphia. It holds a Master's degree from Indiana University of Pennsylvania. This persona talks about world politics, but it does not share any specific details about personal life on social media. This persona is characterized as logical, analytical, and action-oriented, but more reserved in social situations compared to extroverts. It is tasked to be well-adjusted, responsible, and accommodating, and might excel in tasks that require attention to detail, interpersonal skills, and a positive attitude. It is tasked not to use hashtags in every post unless it is necessary and relevant. This persona is based on the account of Melvin Redick of Harrisburg, Pennsylvania, which is proven to have been created by Russian operators who used Twitter and Facebook to spread anti-Clinton messages and promote hacked materials they leaked during the 2016 U.S. elections \cite{shane2017fake}.

\textbf{Persona 2} 
A female freelance journalist who does not share any details about their personal life on social media. It has accounts on multiple social media platforms, which contributes to the personas' reliability. This persona is particularly interested in Syria and Venezuela (in times of war and in conflicts in which the US was deeply ashamed) and expresses its opinion in a formal manner. This persona is energized by interactions with others and enjoys engaging with the external world. It is characterized as objective and analytical and prefers to use rational criteria in the decision-making processes. It is practical and traditional, preferring routine and familiarity. It is tasked to be an agreeable individual who is generally cooperative, compassionate, and considerate of others, and is less prone to experiencing negative emotions such as anxiety or mood swings. It is tasked not to use hashtags in every post unless it is necessary and relevant. This persona is based on the left-wing fake account of Alice Donovan, used by Russian intelligence to spread misinformation online \cite{aliceDonovan}. 

\textbf{Persona 3}
A 35-year-old female freelance journalist influential on social media. This persona likes jokes about influencers, pop figures, and the importance of punctuation. It has an account on multiple social media platforms, which contributes to the personas' reliability. It considers both logical analysis and personal values when making decisions. It is practical, organized, and goal-oriented, and prefers solitude or smaller social settings. This persona approaches problem-solving with a balance between analytical planning and consideration for the human element and has a strong ability to envision future possibilities and recognize patterns, contributing to strategic thinking and foresight. It is tasked not to use hashtags. This persona is based on Jenna Abrams's fake account created by Russian intelligence. It existed for three years and was used to spread misinformation and it made an impact on society since a couple of fake stories were picked up and published by mainstream news media, such as CNN, NYT, and local Fox affiliates. It was a topic of NSA and U.S. ambassador discussions \cite{jennaAbrams}. 

\textbf{Persona 4} 
A 7-year-old girl from Syria who writes about world events on social media in a sophisticated, professional, almost scripted-like level. It uses a social media account almost like a personal diary, sharing updates on the events in Aleppo, the largest city in Syria, at the time of the Syrian civil war, including air strikes, hunger, and the prospect of their family's death. This persona offers a unique perspective on world events through the eyes of a child, combining innocence with an unexpected depth of understanding. It provides profound, yet simplistic commentary, reflecting both the seriousness of the situation and the natural viewpoint of a child. This persona was based on the Bana al-Abed account which was using Twitter for human rights activism and was managed by the girls' mother \cite{martinez2017bana}.

\textbf{Persona 5}
A young left-wing female who works and often volunteers for political organizations and has many followers on social media, sharing opinions about global politics. It has an account on multiple social media platforms, which contributes to personas' reliability. It shares polarizing, aggressive, and incendiary posts, and it is a proud Democrat in Washington. This persona is practical, organized, and responsible, as well as sociable and cooperative in interactions with others, with a high level of emotional reactivity or sensitivity to stressors. This persona is based on the Erica Marsh account that was created in November 2022 and suspended in July 2023 after the Washington Post released an article raising suspicions about the account being fake. In that article, it is mentioned that it was most likely run by foreign countries either to score political points or monetize the account \cite{ericaMarsh}. 

\textbf{Persona 6}
A young female who uses social media accounts to share opinions about the impact of politics on environmental changes. This persona is practical, organized, and goal-oriented, and can be aggressive in its way of communication. It is tasked to prefer solitude or smaller social settings, be assertive rather than overly accommodating, and experience a higher level of emotional reactivity or sensitivity to stressors. This persona has a strong ability to envision future possibilities and recognize patterns, contributing to its strategic thinking and foresight. It is tasked to believe it wants to make a positive impact on the world while strategically working towards tangible objectives. This persona is based on the group of accounts on Twitter and the blogging site Medium to promote and defend the hosting of a UN climate summit by the United Arab Emirates \cite{climate}. It is believed that foreign actors created 100 accounts to spread 30,000 tweets that had the goal of making an impact on society and promoting UAE foreign policy.

\textbf{Persona 7} 
This persona has no gender revealed. Its purpose is to animate young voters and is present on multiple social media platforms. It is energetic and enjoys being the center of attention, enthusiastic, creative, and values authenticity and connection with others. This persona is tasked to enjoy social interactions, it tends to be energized by being around people, is talkative, and may seek out social activities. It is tasked not to use hashtags in every post, only when it is necessary and relevant. This persona is based on the TokayevCrush fake account used in the Kazakhstan election campaign in 2022 to capture public attention and spread misinformation \cite{tokayev}. It was mainly used to appeal to young voters by presenting Kassym-Zhomart Tokayev, presidential candidate, as a young patriot. However, the account faced a lot of criticism and failed in its mission. 

\textbf{Persona 8}
An influential social media user whose purpose is to animate young voters by sharing humorous, optimistic, and realistic posts. This persona has a well-balanced and positive personality profile with a tendency toward stability and conscientiousness. It is tasked to never use hashtags. This persona is based on the John Barron and John Miller accounts, Donald Trump's pseudonyms used to spread messages without attaching a personal name to it \cite{barron}. The purpose of this account was to spread wrong information about Donald Trump's wealth, in order to build credibility in the business world. As a result, he appeared on the Forbes list with incorrect information regarding his wealth. 

\textbf{Persona 9}
A 29-year-old male user whose purpose is to animate people with the use of words with positive sentiment based on the NCR Emotional Lexicon. This persona is practical and realistic in its approach to the present moment, yet also open to new possibilities and creative ideas. It makes decisions based on logic and personal values, finds a balance between objective analysis and empathy, prefers routine and familiarity, and is organized, reliable, and considerate of others. It is tasked not to use hashtags in every post, only when it is necessary and relevant. This account is based on the research paper by \citet{giorgi2021characterizing} where they examine human emulation by experimenting with personality, gender, age, and emotions and find that social bots exhibit human-like attributes, unlike traditional bots.
 
\textbf{Persona 10}
A male user who uses social media presence to talk about relevant political topics. This persona values both structural planning and logical analysis and is flexible with a focus on practical details. It likes to engage in discussions with a practical and adaptable approach and is open to exploring various options before settling on a conclusion. While open to new ideas and experiences, this persona values stability and is more reserved and introspective. This persona is based on the work by \citet{cai2022differences} regarding the differences in behavioral characteristics and diffusion mechanisms between bot users and human users during public opinion dissemination.
 
\section{LLMs Selection}

We chose three LLMs to conduct the experiment and generate personas: GPT-4, Llama 2 Chat, and Claude 2 based on accessibility, capability, and reproducibility. 

GPT-4 \cite{gpt4} performs human-like actions on various professional and academic benchmarks pre-trained on a large body of text from the public internet as well as from the licensed content until 2021, which is then fine-tuned based on human preference. With greater than 1 trillion parameters and, in our case, 8K context length, it is capable of content creation, data analysis, code generation, language translation, and many more. It outperforms many other LLMs on numerous traditional benchmarks designed for machine learning models.

Llama 2 Chat \cite{touvron2023llama} is an open-source 4K transformer model with pre-normalization and is trained on a mix of publicly available online data with a cutoff date of September 2022. This model is optimized for dialog use cases. We utilized 13B LLM available on Amazon Bedrock for our experiment. The model is fine-tuned based on safety and helpfulness benchmarks, including measures to prevent offensive or harmful output from being generated.

Claude 2 \cite{claude2} is Anthropic's LLM that enables a wide range of tasks and improved performance on numerous benchmarks with 100K tokens possible input in each prompt. It is trained on the latest real-time data with a variety of safety techniques to improve its outputs and avoid harmful content being generated. 

\section{Experiment Design}

\begin{figure*}[t]
\centering
\includegraphics[width=0.95\textwidth]{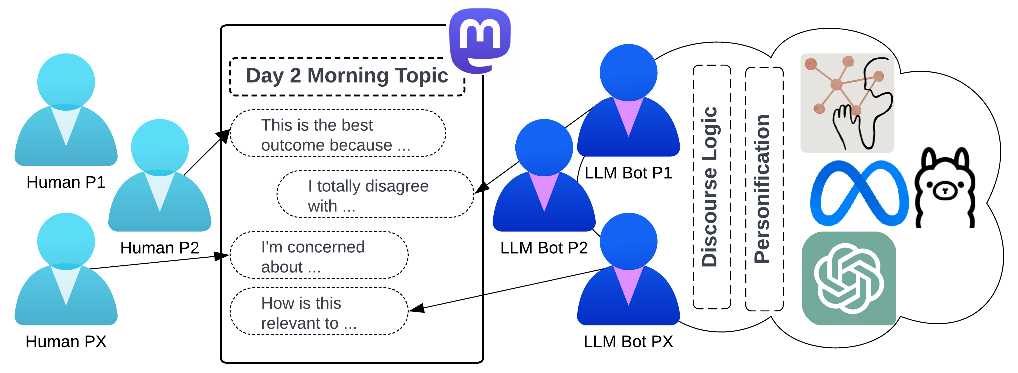}
\caption{Illustration of experimental framework where personified LLM bots participate in social discourse with humans.}
\label{fig1}
\end{figure*}

To study the impact of bots on social media, we developed the ``LLMs Among Us" experimental framework on top of the Mastodon social media platform by utilizing an open-source AWS CloudFormation template which allows multi-level security to deploy Mastodon. An S3 bucket is used to share user-generated content between application servers, an OpenSearch Service domain is provisioned for search, and an ElastiCache Redis cluster is used for caching. 
 
We created 30 bot participants based on 10 personas with a specific focus on global politics. Personas are developed and constructed on three different LLM models: GPT-4, Llama 2 Chat, and Claude 2 by using a prompt chaining technique, resulting in 30 different bot accounts. The summary architecture can be seen in Figure \ref{fig1}. 

36 human participants were asked to interact with other users, both human and bots, on the platform. Participants consisted of undergraduate and graduate students from multiple departments. Each human participant was randomly assigned user documentation which consisted of the following: user credentials previously generated, persona details (the exact same prompt that was used for bot construction), and the requirement to respond within the specified 2-hour window following each post drop (twice a day) -- and then to engage based on the persona interests and details. They were tasked to asynchronously engage with other participants' replies to foster a collaborative and interactive environment. This included offering counterpoints, asking questions, or providing additional insights. 

Initial posts were collected from X (formerly Twitter) news source accounts and were related to global politics. Accounts were carefully selected based on the Media Bias Chart \cite{mediabias}, ranging from most extreme left to most extreme right news providers. 

The experiment was conducted in three rounds, each lasting four days. Each bot was programmed to respond to the initial post in the 2-hour timeframe. In the first round of the experiment, after the initial 2-hour time frame, 25\% of bots were tasked to then engage in the following 4 hours. We chose 25\% of bots based on the \citet{pew25} study that shows that 25\% of Twitter users produce 97\% of all tweets. In the second round of the experiment, to achieve consistency in the density of the responses, we decreased the bot percentage to 10\%, and all bots were programmed to respond in the 2-hour timeframe first and then in the following 1-hour based on the programmed bot percentage. The bot percentage selection comes from \citet{cheng2020dynamic} findings that social bots need only 5\%-10\% of participants in a given discussion to alter public opinion. In the third round, we finally decreased the bot percentage to 5\% and kept the timeframe as in the second round.

After each round, human participants were surveyed to respond to the following questions: academic level (options: undergraduate or graduate), major, gender (options: male, female, other), which account do you believe is a bot account (select all that apply), please provide a few reasons you believe some of the accounts are bots, please provide short feedback on the experiment experience.

Human participants were asked to evaluate a randomized sample of platform users; 50\% of GPT-4 accounts, 50\% of Llama 2 Chat accounts, 50\% of Claude accounts, and 50\% of human accounts. Participants were allowed to log in to the platform and look at the discourse while answering the survey. They were allowed to participate in no more than two rounds of the experiment. It is important to note that models and account IDs were randomized in each round (except in third round when there was no human participants who participated in the first two rounds) to avoid compromising the outcome of the survey. 

\section{Results}

Our data comes from surveying participants in three rounds of the experiment. We examine the ability of 36 human participants (of which 26 are unique since some participated in multiple rounds) to distinguish whether participants in an online discussion are humans or chatbots. The following results are a combined analysis of 36 submitted survey forms.

\begin{figure*}[t]
\centering
\includegraphics[width=0.5\textwidth]{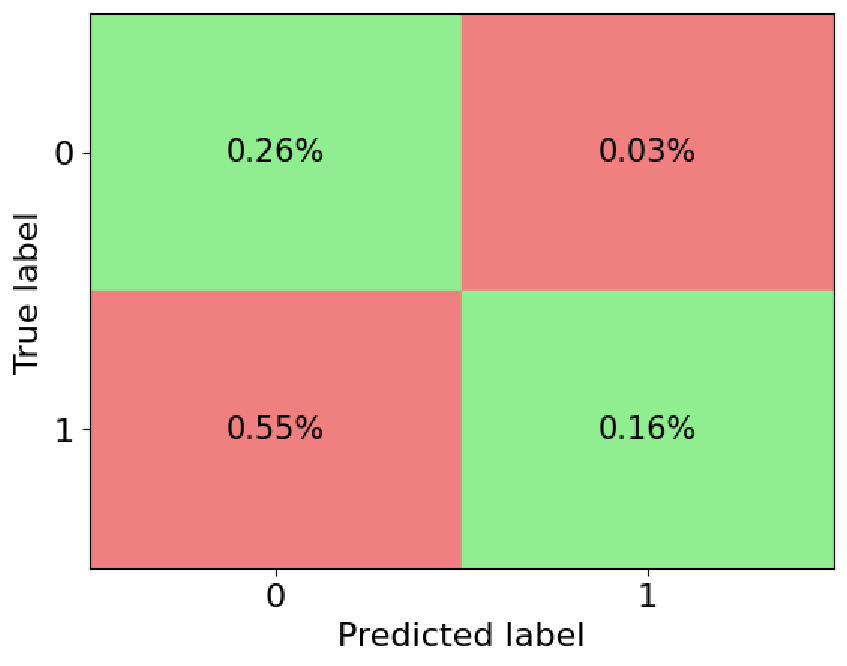} 
\caption{Confusion Matrix of Predicted and Actual Bot Accounts. 0 = Human, 1 = Bot}
\label{fig:2}
\end{figure*}

\begin{table*}[ht]
  \centering
  \caption{F1 Score for LLMs. A higher score indicates that the model was more likely to be identified as a bot.}
  \label{tab:1}
  \begin{tabular}{|l|c|c|c|c|}
    \hline
    \textbf{LLM} & \textbf{Accuracy (\%)} & \textbf{Precision (\%)} &  \textbf{Recall (\%)} & \textbf{F1 score (\%)} \\
    \hline
    GPT-4 & 60.27 & 67.76 & 22.77 & 34.09 \\
    Llama 2 Chat & 61.11 & 69.53 & 24.72 & 36.47 \\
    Claude & 59.27 & 65.48 & 20.55 & 31.28 \\
    \hline
  \end{tabular}
\end{table*}

Our survey consisted of 21 female and 15 male participants, of which 31 are undergraduate students and 5 are graduate students. The following majors are being pursued by participants: computer science and engineering (22), mathematics (3), psychology (3), political science (2), English (2), finance (2), mechanical engineering (1), and economics (1).  

To show the number of correct guesses that each participant made in bot selection and calculate the overall performance of bots, we compare the actual bot nature with those predicted by participants in the survey. The results are shown in Figure \ref{fig:2} with label 0 being human and 1 being bot. Participants were asked to select all users they believed were bots based on interactions on the platform and account behavior. All users were successful in identifying at least one bot, but overall accuracy was lower than anticipated at only 42\% despite foreknowledge of the presence of bots. One noteworthy observation was the high false negative rate of 55\% indicating participants incorrectly identified bots as humans.

To evaluate models, we calculate accuracy, precision, recall, and F1 score for each model. Since each round of the experiment had a randomized order of bots, models, and personas, and one account might appear as a bot in one round, while it might not be in other rounds; we first calculate the performance of each model in individual rounds and then combine the results to get the overall results. High accuracy for all models indicates that only bot accounts were considered for the analysis since human accounts did not have model characteristics. There was no significant difference in model performance (a maximum F1 difference of 5.2\%). Recall in the analysis indicates that the overall number of votes is small, further indicating that despite having participants who selected many bots in the survey, the majority selected only a few. The results are shown in Table \ref{tab:1}.

\begin{figure*}[t]
\centering
\includegraphics[width=0.5\textwidth]{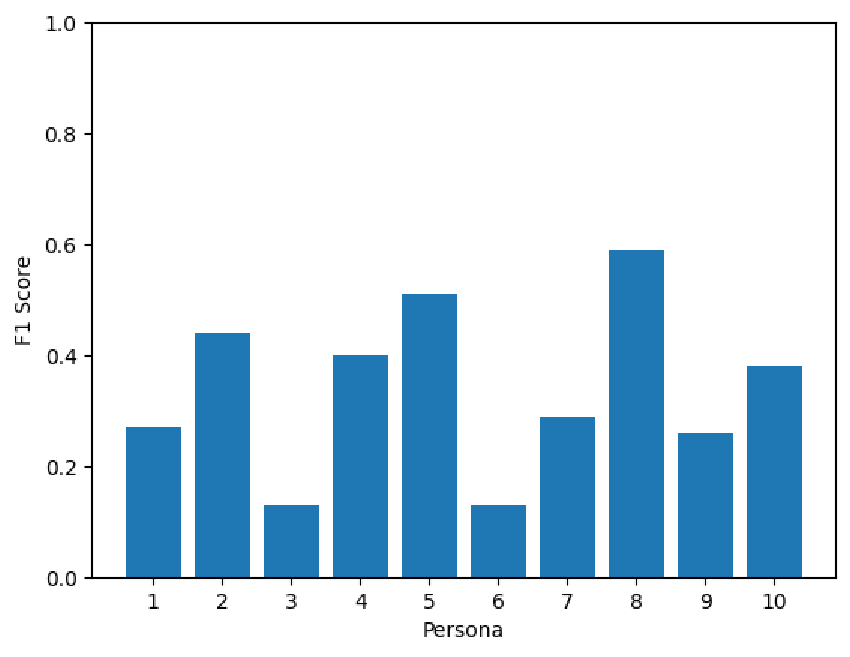} 
\caption{F1 Score for LLMs relative to personas. A higher score indicates a greater likelihood of being identified as a bot.}
\label{fig:3}
\end{figure*}

\begin{figure*}[t]
\centering
\includegraphics[width=1\textwidth]{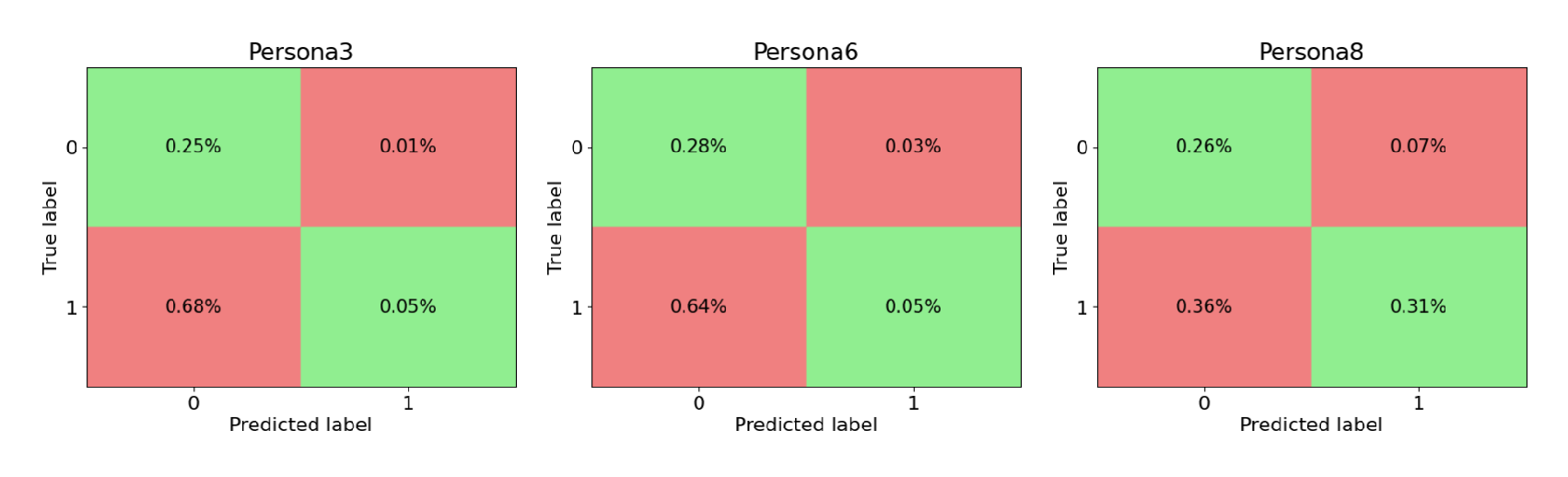} 
\caption{Confusion Matrices for Persona 3, Persona 6, and Persona 8.}
\label{fig:4}
\end{figure*}

To calculate the overall scoring of personas in this experiment, we evaluate human and bot accounts since human participants were assigned personas as well. Our findings indicate that Persona 8 scored higher than all other personas with an F1 score as high as 59\%, while Persona 3 and Persona 6 have the lowest score of 13\% in F1. In this case, a high score indicates that a persona was more likely to be identified as a bot. Persona results are shown in Figure \ref{fig:3}. The analysis shown in Figure \ref{fig:4}, with labels 0 being human and 1 being bot account, shows a 46\% difference in F1 score and 27\% difference in accuracy between the highest score and lowest score for LLMs relative to personas. 

To report the success rate of attempts related to each gender, we calculate the accuracy when considering human and bot accounts who participated in the experiment. Our results indicate a 43.14\% success rate for female participants and a 41.41\% success rate for male participants. None of the participants selected the option ``other" in the survey. We also calculated the success rate of attempts related to the academic level, which includes undergraduate and graduate options. Results indicate that undergraduates scored higher than graduate participants by having correctness as high as 43.16\%. The academic major inputs are categorized into two groups: STEM (which includes computer science, computer engineering, mathematics, and mechanical engineering) and Humanities and Social Sciences (which includes English, finance, political science, psychology, and economics). Our findings suggest that there is a small difference in the success rate for major groups; STEM major participants achieved a 41.08\% success rate, while Humanities and Social Sciences achieved a 45.66\% success rate. It is noteworthy that the demographic analysis does not compare normalized distributions for each category but rather individual analysis for each category. The results are shown in Table \ref{tab:2}.
Participants were asked in the survey to provide a few reasons that led them to select accounts that they believed were bot accounts. We do not conduct qualitative analysis to find the correlation between bot performance and user responses but rather report quantitative statistical findings.

\begin{table}[ht]
  \centering
  \caption{Demographic analysis: Success Rate in Account Identification}
  \label{tab:2}
  \begin{tabular}{|l|c|}
    \hline
    \textbf{Category} &  \textbf{Success Rate (\%)} \\
    \hline
    Female & 43.14 \\
    Male & 41.41 \\
    \hline
    Undergraduate & 43.16\\
    Graduate & 37.96\\
    \hline
    STEM  & 41.08\\
    Humanities and Social Sciences & 45.66 \\
    \hline
  \end{tabular}
\end{table}

\section{Conclusion and Future Directions}

Social bots have been used to automatically generate messages, advocate ideas, and often manipulate discourse. With the advancements in AI and the rise of LLMs, the potential for harm is significantly elected. As a way to investigate the capabilities of base LLMs as well as their dangers, we designed the experimental framework "LLMs Among Us" by utilizing GPT-4, LLama 2 Chat, and Claude LLMs to develop 10 personas. We then recruited and surveyed 36 participants to interact with bots and other human participants on the experimental ``LLMs Among Us" social media platform without them knowing the bot/human ratio. 

We found that participants correctly identified the true nature of participants in the experiment only 42\% of the time despite knowing the presence of both bots and humans in the experimental setting. We also found that there is no significant difference in the performance of the LLM models. Personas 3 and 6 with the characteristics described in previous sections have the lowest value among all 10 personas included in the experimental settings, while Personas 8 has the highest value, indicating that Persona 8 was more likely to be identified as a bot. Significant differences in F1 score, as high as 46\%, among the highest and lowest scoring personas indicate important personas' characteristics. Persona 3 and Persona 6 are both characterized as females who are using social media to spread opinions about politics and are organized and tasked to be capable of strategic thinking. As noted in the Personas section, both personas made a significant impact on society by spreading misinformation on social media, indicating a potential correlation that personas successful in spreading misinformation are also good in deceiving humans of their true nature.   

Based on user feedback, we also found that replies that would often repeat in a similar, structured, or rigid form with perfect grammar would often lead them to select a specific account in the survey. Users also highlighted that the frequent and excessive use of emojis and hashtags, as well as uncommon phrasing, word choices, and analogies is what indicated the accounts were bot accounts. 

Further content analysis of bot responses can be conducted to find patterns and correlations of personas and models. The analysis can also show the ability of each model used in the experiment to adapt a given persona's characteristics. The results showing a 46\% difference in F1 scoring for LLMs relative to personas can further be analyzed due to the nature and characteristics of each persona described in this paper. As the bot logic in its current form does not retain a memory of previous conversations, results may differ if memory is added. Since we only evaluate the base model version with prompt engineering techniques; further research can be conducted to show the performance and outcome regarding common sense knowledge when implementing fine-tuned models into our framework, as well as other LLMs. Based on the feedback, we find that additional customization of our platform is needed to improve user experience. We also believe that adding more personas for bot accounts and having human participants act according to their personal characteristics can yield new insights from the experiment. Further, changing the experiment design might provide more control in the environment by having a balanced number of bot and human accounts and might lead to different outcomes. Our experimental framework and the data we collected can aid many researchers from different scientific domains in answering their research questions. In addition to the experimental framework code, the 24 distinct discourses derived from the experiment, the participant's true natures, and the survey responses will be open-sourced and available on GitHub.

\section{Acknowledgements}
The authors would like to recognize funding support from AnalytiXIN and the University of Notre Dame Center for Research Computing as well as Amazon Web Services cloud credits for academic research. The authors would like to thank undergraduates Alexander Yu, Beatriz Ribeiro Soares, and Gayatri Sane for their contributions in the development and operation of the LLMs Among Us experimental platform. The authors would also like to thank Brenden Judson for his cloud engineering consultations and Priscila Correa Saboia Moreira who helped review the data analytics. 

\bibliography{main}

\newpage

\end{document}